\documentclass[prb,twocolumn,superscriptaddress,floatfix,showpacs]{revtex4}
\usepackage{graphicx,amsfonts,amssymb,amsmath, hyperref}

\newif\ifhyper
\hypertrue
\ifhyper
\hypersetup{
   citecolor = {green},
   colorlinks = {true}, 
   urlcolor = {blue} 
}
\fi

\newcommand{\beq}{\begin{equation}}
\newcommand{\eeq}{\end{equation}}
\newcommand{\beqa}{\begin{eqnarray}}
\newcommand{\eeqa}{\end{eqnarray}}
\newcommand{\ket} [1] {\vert #1 \rangle}

\def\ket#1{\vert#1\rangle}

\def\ipr#1#2{\langle#1\vert#2\rangle}

\def\Longarrow{\protect\@lra}
\def\@lra{\relbar\joinrel\relbar\joinrel\relbar\joinrel%
          \relbar\joinrel\rightarrow}

\begin{document}

\title{Phase diagram of the SO($n$) bilinear-biquadratic chain from many-body entanglement}

\author{Rom\'an Or\'us}
\affiliation{School of Mathematics and Physics, The University of Queensland, QLD 4072, Australia}
\affiliation{Max-Planck-Institut f\"ur Quantenoptik, Hans-Kopfermann-Str. 1, 85748 Garching, Germany}

\author{Tzu-Chieh Wei}
\affiliation{Department of Physics and Astronomy, University of
British Columbia, Vancouver, BC V6T 1Z1, Canada} \affiliation{C. N.
Yang Institute for Theoretical Physics, State University of New York
at Stony Brook, NY 11794-3840, USA}

\author{Hong-Hao Tu}
\affiliation{Max-Planck-Institut f\"ur Quantenoptik, Hans-Kopfermann-Str. 1, 85748
Garching, Germany}

\begin{abstract}
Here we investigate the phase diagram of the SO($n$) bilinear-biquadratic quantum spin chain by studying the global quantum correlations of the ground state. We consider the cases of $n=3,4$ and $5$ and focus on the geometric entanglement in the thermodynamic limit. Apart from capturing all the known phase transitions for these cases, our analysis shows a number of novel distinctive behaviors in the phase diagrams which we conjecture to be general and valid for arbitrary $n$. In particular, we provide an intuitive argument in favor of an infinite entanglement length in the system at a purely-biquadratic point. Our results are also compared to other methods, such as fidelity diagrams.

\end{abstract}


\maketitle

\section{Introduction} The recent ability to experimentally manipulate ultracold atoms in optical lattices has opened the possibility to simulate a number of strongly-correlated systems \cite{cold-1, cold-2}. In fact, it is now conceivable to recreate the physics of some quantum spin systems of relevance in condensed matter physics \cite{spin}. This great advance in quantum simulation has opened the possibility to understand better many relevant phases of matter, such as the celebrated Haldane phase of quantum spin chains. First encountered analytically at a so-called Affleck-Kennedy-Lieb-Tasaki (AKLT) point of the spin-1 antiferromagnetic Heisenberg model with a bilinear-biquadratic interaction \cite{aklt-1,aklt-2}, this phase has become a landmark in the study of strongly correlated quantum systems in one spatial dimension.

The existence of a Haldane phase in integer quantum spin chains motivated the study of generalizations of the spin models with SU(2) symmetry. For instance, generalized AKLT models with a Valence Bond Solid structure have been proposed, which promote SU(2) symmetry to some larger Lie groups such as SU($n$) \cite{genaklt-1, genaklt-2, genaklt-3}, Sp($2n$) \cite{genaklt-4} and SO($n$) \cite{son-1, son-2} together with supersymmetric versions \cite{genaklt-5,genaklt-6}. Many exciting results also have been found in more general quantum spin chains with such symmetries, including the existence of Haldane-like phases.

\begin{widetext}

\begin{figure}[h]
\includegraphics[width=1\textwidth]{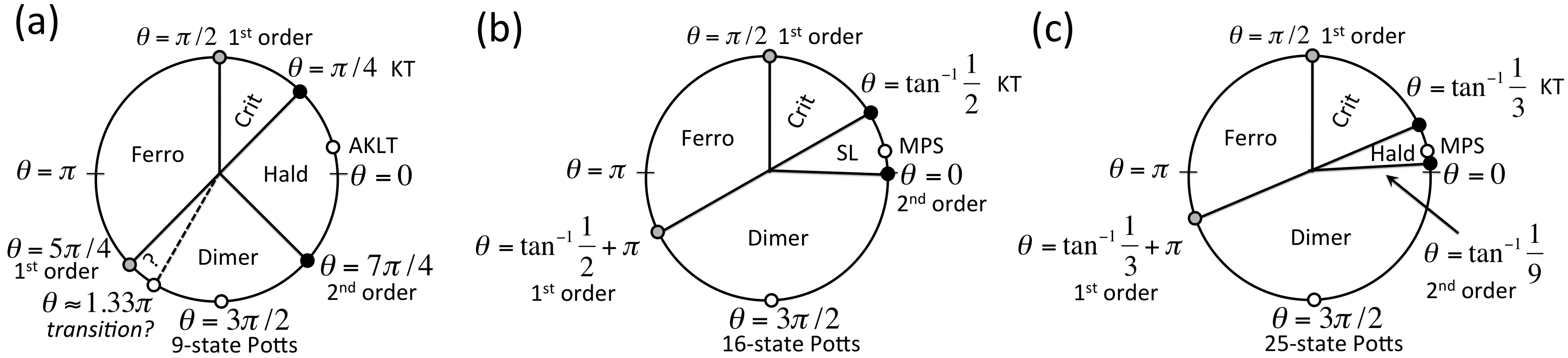}
\caption{Phase diagram of the SO($n$) bilinear biquadratic chain for (a) SO($3$), (b) SO($4$) and (c) SO($5$). Black and grey dots indicate respectively known continuous and discontinuous transitions, whereas white dots indicate other interesting points of the system. The white dot close to $\theta = 0$ corresponds to the MPS point $\tan{\theta} = 1/n$ in all cases (AKLT for $n=3$).} \label{figbbdiag}
\end{figure}

\end{widetext}

Here we wish to focus on the case of SO($n$)-symmetric quantum spin chains. In particular, our aim is to further contribute to the study and understanding of the different phases present in the SO($n$) bilinear-biquadratic quantum spin chain
\begin{equation}
H=\cos \theta \sum_{j}\sum_{a<b}L_{j}^{ab}L_{j+1}^{ab}+\sin \theta
\sum_{j}(\sum_{a<b}L_{j}^{ab}L_{j+1}^{ab})^{2}  \label{bb}
\end{equation}
In the above expression
$L^{ab}$ ($1\leq a<b\leq n$) are the generators of SO($n$) in the
$n$-dimensional vector representation, with the Casimir operator
normalized as $\sum_{a<b}(L^{ab})^{2}=n-1$, and $\theta \in
[0,2\pi)$ is the Hamiltonian parameter. Despite its simplicity, the
model is known to have a very complex and rich phase diagram
offering a wide variety of phases \cite{son-1, son-3, son-4} (see
Fig.~\ref{figbbdiag}). In fact, this phase diagram is not totally
understood yet. Even in the simple case of $n=3$, where the model
corresponds to the well-known spin-1 bilinear-biquadratic quantum
spin chain, there is controversy about the presence or not of a
spin-nematic phase \cite{nem-1, nem-2, nem-3, nem-4, nem-5, nem-6,
nem-7, nemtr-1, nemtr-2} near $\theta = 5\pi/4$ \cite{chu}.

In this paper we investigate the phase diagram of the above model by focusing on a \emph{global} property of the ground state in the thermodynamic limit, namely, the \emph{geometric entanglement} (GE) \cite{ge}. This quantity has proven useful in the study of several quantum many-body systems and their associated quantum phase transitions without the need of order parameters \cite{geqpt-1, geqpt-2, geqpt-3, geqpt-4, geqpt-5, geqpt-6, geqpt-7, geqpt-8, geqpt-9, geqpt-10, geqpt-11, geelusive}. Here, we show that for the model in Eq.~(\ref{bb}) the GE shows a rich behavior as compared to other quantities such as the mutual information \cite{mutual}, fidelity susceptibility \cite{fs}, R\'enyi entropies and correlation functions, and is compatible with calculations of the fidelity diagram \cite{fdqpt-1, fdqpt-2, fdqpt-3, fdqpt-4, fdqpt-5, fdqpt-6, fdqpt-7, fdqpt-8}. More specifically, we consider numerically the cases of $n=3,4$ and $5$. As we shall see, our analysis is able to capture distinctive behaviors which we conjecture to be general and valid for arbitrary $n$. For instance, an anomaly in the GE is observed at the purely bilinear point $\theta = 3 \pi /2$, where the model is dual to the $n^2$-state Potts model and its ground state energy and gap can be obtained exactly \cite{bethe-1, bethe-2, bethe-3}. This motivates us to give an intuitive argument in favor of an infinite entanglement length and finite correlation length at this point \cite{localizable}.

This paper is organized as follows: in Sec. II we explain the basic phase diagram of the model for SO(3), and also on the known properties of the phase diagram for SO($n$). In Sec. III we comment briefly on the methods we have used in our study, including the entanglement measure and the numerical technique. Sec. IV contains our results for the GE for SO(3), SO(4) and SO(5), together with a discussion about these results and the behavior for arbitrary $n$. In this section we also compare our results to those obtained by alternative approaches. Finally, in Sec. V we briefly summarize our conclusions. For completeness, we include in Appendix A the calculation of the fidelity diagram for the SO(3) case.

\section{Basic phase diagram}
\subsection{SO(3)}
Let us start by quickly reminding the phase diagram of the model for the case of SO($3$), see Fig.~\ref{figbbdiag}(a). In the region $\theta \in (7\pi/4,\pi/4)$ the system favors a Haldane gapped phase, and at $\theta = 0.1024 \pi$ ($\tan{\theta} = 1/3$) the system corresponds to the AKLT model with an exact valence bond ground state \cite{aklt-1, aklt-2}. At $\theta=\pi/4$ there is a continuous phase transition of the Kosterlitz-Thouless (KT) type to a critical trimerized phase with $k = \pm 2\pi/3$ spin quadrupolar correlations, which extends through the region $\theta \in (\pi/4, \pi/2)$. This phase is separated from a gapless ferromagnetic phase in $\theta \in (\pi/2, 5\pi/4)$ at a first order transition point $\theta = \pi/2$. At $\theta = 5\pi/4$ there is another first order transition to a dimerized phase that extends through $\theta \in (5\pi/4, 7\pi/4)$. It has been suggested that a spin nematic phase appears at the region $\theta \in (5 \pi/4, 1.33 \pi)$ with a generalized KT transition around $\theta \approx 1.33 \pi$ \cite{nemtr-1, nemtr-2}. The existence or not of this spin nematic phase is still a matter of discussion \cite{nem-1, nem-2, nem-3, nem-4, nem-5, nem-6, nem-7, nemtr-1, nemtr-2}. Also, many of the properties of the system can be obtained exactly at the purely biquadratic point $\theta = 3 \pi/2$, where it is dual to the 9-state Potts model \cite{bethe-1, bethe-2, bethe-3}. The quantum phase transition at $\theta = 7 \pi/4$ between the dimerized and Haldane phases is of second order. Also, the system is exactly solvable at the transition points $\theta = \pi/4, \pi/2, 5 \pi/4$ and $7 \pi/4$.

\subsection{SO($n$)}
The case of general SO($n$) is similar, see
Fig.~\ref{figbbdiag}(b,c) for $n=4$ and $5$. In general, the phase
that extends through the region $\theta \in
(\tan^{-1}{\frac{(n-4)}{(n-2)^2}}, \tan^{-1}\frac{1}{(n-2)})$ always
contains an MPS point at $\tan{\theta} = 1/n$. This phase is
Haldane-like for $n=3,5$ and a non-Haldane spin liquid for $n=4$
\cite{mpspoint-1, mpspoint-2}. Field theory analysis seems to
indicate that there is a KT transition at $\theta =
\tan^{-1}\frac{1}{(n-2)}$ towards a critical phase \cite{field}. At
$\theta = \pi/2$ there is a first order transition towards a phase
with ferromagnetic order. This phase extends up to $\theta =
\tan^{-1}\frac{1}{(n-2)} + \pi$, where there is a transition to a
dimerized phase. Again, at the purely-biquadratic point $\theta =
3\pi/2$ the model is dual to the $n^2$-state Potts
model~\cite{bethe-1, bethe-2, bethe-3}. We remark that the  Heisenberg 
point $\theta=0$ does not necessarily reside in the same phase as
the MPS point. For details see e.g. Ref.\cite{son-1, son-3, son-4}.

\section{Methods}
\subsection{Geometric entanglement}
Here we choose to explore the details of the above phase diagrams
for $n=3,4$ and $5$ using the geometric entanglement per site
$\mathcal{E}$ (which we refer to as GE) \cite{ge, geqpt-1, geqpt-2,
geqpt-3, geqpt-4, geqpt-5, geqpt-6, geqpt-7, geqpt-8, geqpt-9,
geqpt-10, geqpt-11, geelusive}. Let us briefly remind the basics of
this quantity. Assume we are given a quantum state $\ket{\Psi}$ of
$N$ parties belonging to a Hilbert space $\mathcal{H} =
\bigotimes_{r=1}^N \mathbb{V}^{[r]}$, where $\mathbb{V}^{[r]}$ is
the Hilbert space of party $r$. We now consider the closest
normalized product state of the parties to $\ket{\Psi}$. By
``closest" we mean the normalized product state $\ket{\Phi} =
\ket{\phi^{[1]}}\otimes \ket{\phi^{[2]}} \otimes \cdots \otimes
\ket{\phi^{[N]}}$ that minimizes the squared distance $|| \ket{\Phi}
- \ket{\Psi}||^2$ between $\ket{\Phi}$ and $\ket{\Psi}$ or,
equivalently, maximizes the absolute value of their overlap,
$\Lambda_{\max}({\Psi})\equiv \max_{\Phi}|\ipr{\Phi}{\Psi}|$. This
closest product state approximation to $\ket{\Psi}$ allows us to
quantify its entanglement via the extensive quantity
$E({\Psi})\equiv-\log\Lambda^2_{\max}(\Psi)$, where we have taken
the natural logarithm. As demanded, $E({\Psi})$ is zero if
$\ket{\Psi}$ is unentangled. We can also define the thermodynamic
quantity ${\cal E}$ and its finite-size version ${\cal E}_N$ as
\begin{equation}
{\cal E}\equiv\lim_{N\to\infty}{\cal E}_{N}, \ \ {\cal E}_{N}\equiv
{N}^{-1}E(\Psi).
\label{ene}
\end{equation}
The quantity ${\cal E}$ in the above equation defines the {\it global
geometric entanglement per site}, or {\it density of global geometric
entanglement}. This will be the quantity of interest in this paper. In the past, this quantity has been used in the study of quantum phase transitions \cite{geqpt-1, geqpt-2, geqpt-3, geqpt-4, geqpt-5, geqpt-6, geqpt-7, geqpt-8, geqpt-9, geqpt-10, geqpt-11, geelusive}, quantification of entanglement as a resource for quantum computation \cite{qua}, local state discrimination \cite{loc} and, more recently, it has also been directly measured experimentally \cite{exp}.

\subsection{MPS and iTEBD}
Our results have been obtained by approximating the ground state of
the model in Eq.~(\ref{bb}) by a matrix product state (MPS) in the
thermodynamic limit \cite{itebd-1, itebd-2}. As is well known, MPSs
offer an accurate description of quantum states of 1D quantum
many-body systems, specially for gapped systems. Let us be more
specific: for a system of size $N$ with periodic boundary
conditions, these are states defined as \beq \ket{\Psi} =
\sum_{i_1,\ldots,i_N} {\rm tr}( A^{[1] i_1} \cdots A^{[N] i_N} )
\ket{i_1, \ldots, i_N}, \label{mps} \eeq where $A^{[m] i_m}$ is a
$\chi \times \chi$ matrix at site $m = 1, \ldots, N$ for each $i_m =
1,\ldots, d_m$, which labels a local basis of the Hilbert space of
dimension $d_m$ at site $m$.  Since our model is an SO(n) spin
chain, we have $d_m = n$ at all sites. Moreover, $\chi$ is a
refinement parameter for the MPS and is called the \emph{bond
dimension}. In practice, MPS offers a good variational family of
states to approximate quantum states of many-body systems, specially
for ground states of Hamiltonians with local interactions in one
spatial dimension. The refinement parameter $\chi$ control the
number of variational parameters in the ansatz. Therefore, the
larger is $\chi$, the more accurate is the variational
approximation.

In our case, we choose to use the iTEBD algorithm \cite{itebd-1,
itebd-2}, an extension of the MPS method to infinite systems, to
obtain approximations to the ground state of our quantum spin chain.
This method makes use of an evolution in imaginary time in order to
update the different matrices of the MPS at each time step. The
particular technical details of the method can be found in
Ref.\cite{itebd-1, itebd-2}.

\begin{figure}
\centerline{\includegraphics[width=9cm,angle=0]{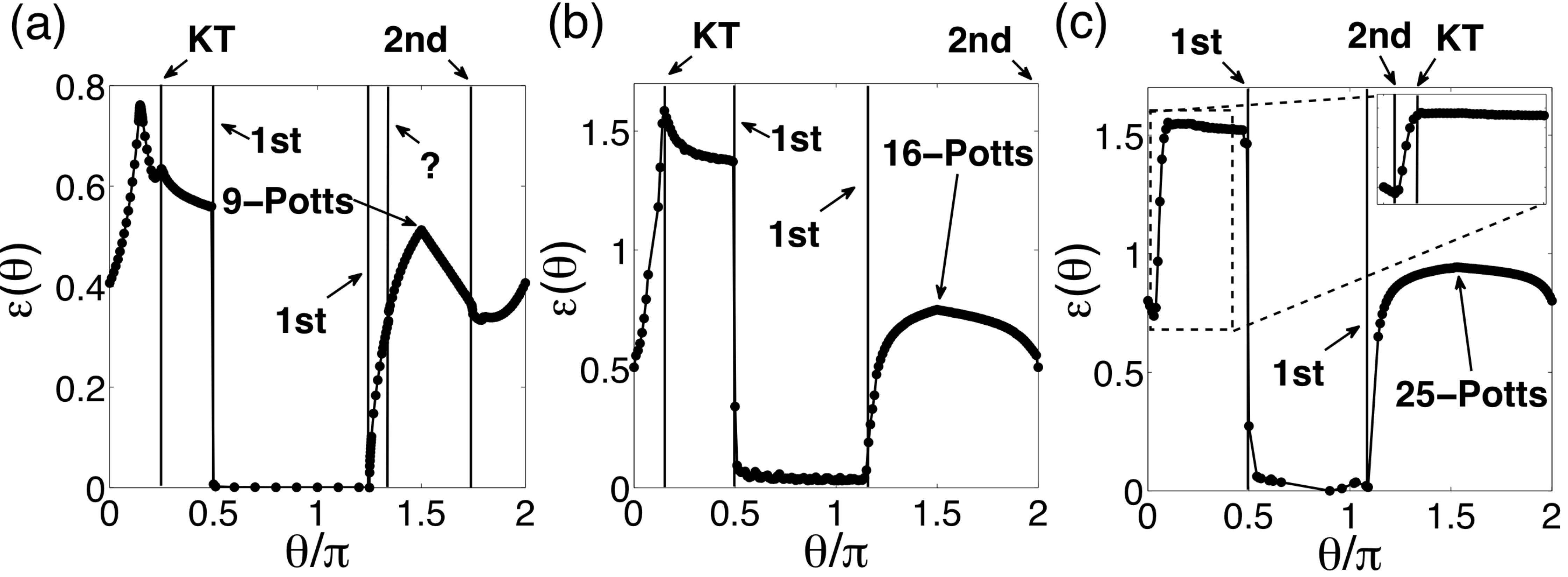}}
\caption{GE of the SO($n$) bilinear-biquadratic model, for (a) SO(3), (b) SO(4) and (c) SO(5).}
\label{fig:GE}
\end{figure}

For our purposes, we find that an MPS with inner bond dimension $\chi$ up to $\sim 120$ is sufficient to achieve convergence in the results \footnote{The GE converges quickly in $\chi$, see \cite{geqpt-1, geqpt-2, geqpt-3, geqpt-4, geqpt-5, geqpt-6, geqpt-7, geqpt-8, geqpt-9, geqpt-10, geqpt-11}.}. Also, the GE is computed from the MPS wave function by using similar techniques as in \cite{geelusive}, but using product state approximations with a periodicity of 6 sites. This helps in avoiding local minima in the optimization process to compute the GE.

\section{Results}
\subsection{GE for SO(3), SO(4) and SO(5)} 
\begin{figure}
\centerline{\includegraphics[width=9cm,angle=0]{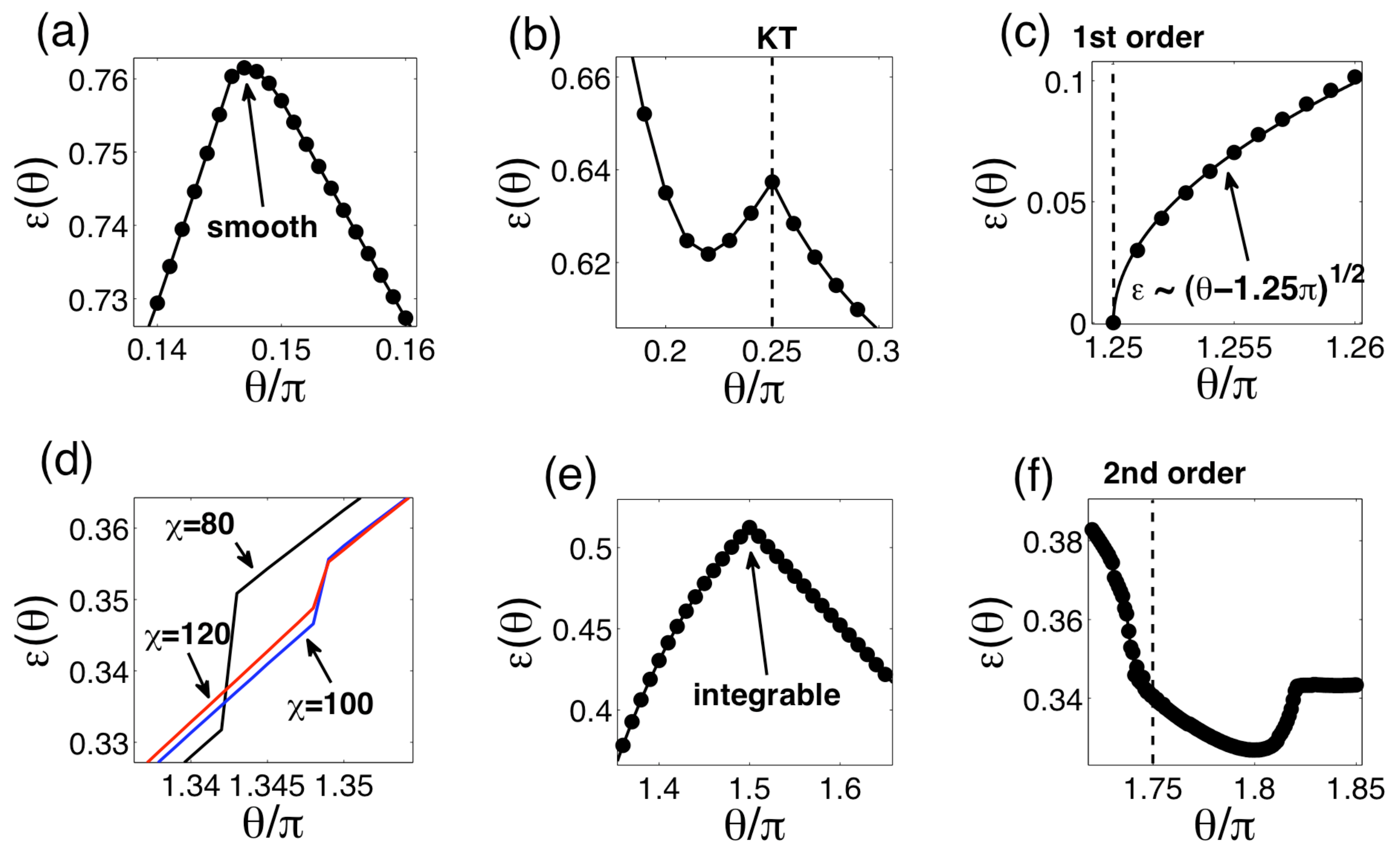}}
\caption{(Color online) Different zooms to the GE diagram from Fig.~\ref{fig:GE}(a): (a) Around $\theta \approx 0.15 \pi$ the GE is smooth (line is a guide to the eye); (b) At the KT transition point $\theta = \pi/4$ the GE shows a cusp (line is a guide to the eye); (c) The GE goes to zero at the first order transition point $\theta = 5\pi/4$, seemingly according to the continuous law $\mathcal{E} \sim \sqrt{(\theta - 5\pi/4)}$ (continuous line); (d) At $\theta \approx 1.34 \pi$ the GE shows a discontinuity, which eventually vanishes as the MPS bond dimension increases (black line is for $\chi=80$, blue line for $\chi=100$, red line for $\chi = 120$, and the position of points has been omitted for the sake of clarity); (e) At the integrable point $\theta = 3 \pi/2$ the GE shows a cusp (line is a guide to the eye); (f) Around the 2nd order transition at $\theta = 7 \pi/4$ the GE shows a smooth change of behavior, compatible with $\mathcal{E} \sim a|\theta_c - \theta|^{\nu}$ with $\nu \sim 1$ (line is a guide to the eye).}
\label{fig:GEZoom}
\end{figure}

Let us now present our results for the GE of the SO(3) model. This is shown in Fig.~\ref{fig:GE}(a) for the whole region $\theta \in [0, 2 \pi)$. Quite remarkably, the GE displays some sort of singularity close to all the known transitions in the system. In particular, at $\theta=\pi/4$ it has a cusp, which we interpret as an indicator of the KT transition between the Haldane and trimerized phases, see Fig.~\ref{fig:GEZoom}(b). This is remarkable, since at a KT transition all the two-body observables as well as their derivatives are analytic, yet the GE is not. A similar cusp in the GE was also found at the KT transition of the XXZ spin-1/2 chain in Ref.~\cite{geelusive}. These cusps in KT transitions are due to a sudden change in the closest product state to the ground state (see Ref.~\cite{geelusive}).

Moreover, the GE has a discontinuity close to $\theta=\pi/2$, which is reminiscent of the first order transition at this point. It is also non-analytic at the first order transition point $\theta=5\pi/4$. Close to this point it arises from zero seemingly according to the power law ${\cal E}\sim \sqrt{\theta-5\pi/4}$, see Fig.~\ref{fig:GEZoom}(c). Notice, though, that this continuous behavior is not incompatible with the presence of a discontinuous transition at $\theta = 5 \pi/4$ \footnote{This is similar to the Ising chain in parallel field \cite{geelusive}.}. We see, thus, that discontinuous transitions can be detected in the GE as (i) discontinuities (as in $\theta=\pi/2$), or (ii) cusps (as in $\theta = 5 \pi/4$). Consequently, in order to distinguish if a given cusp in the GE corresponds to a discontinuous transition of type-(ii) or to a KT transition, one may need to also study e.g. the correlation functions of the system or energy derivatives. In the first case these quantities would be discontinuous across the transition, whereas across the KT transition they would be continuous.

We also observe a peculiar behavior close to the transition at $\theta = 7 \pi /4$, see Fig.~\ref{fig:GEZoom}(f). Specifically, the GE begins to drop sharply near $\theta\approx 1.74 \pi$ and begins to flatten near $\theta \approx 1.75 \pi$. At $\theta\approx 1.81\pi$ the GE sharply arises. Indeed, it is possible to estimate the correlation length exponent $\nu$ from the GE close to the transition at $\theta = 7 \pi /4$, since close to a quantum critical point $\mathcal{E} \sim a|\theta_c - \theta|^{\nu}+b$ \footnote{This follows from Ref.~\cite{geqpt-10}.}. Fitting our results very close to this point is compatible with $\nu = 0.5 \pm 0.1$, in agreement with a mean-field theory estimation $\nu=1/2$. However, if another fit is done slightly away from $\theta = 7 \pi/4$ (from below), we obtain $\nu \sim 1$ instead, as predicted by a conformal field theory approach \cite{affleck}. This cross-over behavior is consistent with the mean-field-like behavior of MPS when close to quantum critical points (see Ref.~\cite{sandvik}).

In addition to the above observations, we see a number of other peculiarities in the GE. For instance, the data between $\theta=3\pi/2$ and $\theta \sim 1.7 \pi$ can be very well fitted by a straight line. We also observe that near $\theta=0.15 \pi$ it attains a maximum. A closer look to this maximum shows that it is smooth, see Fig.~\ref{fig:GEZoom}(a). The presence of this maximum is intriguing as at this point the ground state possesses the highest geometric entanglement in this model, albeit we are not aware of this point being special for any other reason. Notice also that there is no sign of any peculiarity in the GE around the AKLT point at $\theta \approx 0.1024 \pi$. However, we observe two other distinctive behaviors. First, near $\theta \approx1.34\pi$ the GE seems to display a small discontinuity, see Fig.~\ref{fig:GEZoom}(d). As seen in the figure, an analysis for large bond dimension shows that this discontinuity eventually vanishes as the MPS bond dimension increases, and is therefore an artifact of the numerical calculation and not caused by the hypothesized transition towards a spin nematic phase (which some studies estimate around $\theta \approx 1.33 \pi$) \cite{nemtr-1, nemtr-2}. Second, the GE displays a cusp at $\theta=3 \pi/2$, see Fig.~\ref{fig:GEZoom}(e). At this point the system is purely biquadratic and dual to the nine-state Potts model \cite{bethe-1, bethe-2, bethe-3}. Such a cusp looks similar to the one in the elusive transition of the deformed AKLT model, where the entanglement length diverges while the correlation length is finite \cite{geelusive, localizable}. We will further analyze the physics of this point later on.

In the case of SO(4) and SO(5), the behavior of the GE shares many basic features with the previous calculation for SO(3), see Fig.\ref{fig:GE}(b,c). Specifically, we observe: (i) a peak at the KT transition point $\theta = \tan^{-1}\frac{1}{(n-2)}$; (ii) a discontinuity at the first order transition in $\theta = \pi/2$; (iii) a non-analyticity at $\theta = \tan^{-1} \frac{1}{(n-2)} + \pi$ consistent with a discontinuous transition; (iv) a peak at the biquadratic point $\theta = 3\pi/2$; (v) a linear behavior between $\theta=3\pi/2$ and $\theta \sim 1.7 \pi$;  and (vi) a sudden change of behavior at $\theta=\tan^{-1}{\frac{(n-4)}{(n-2)^2}}$ compatible with a continuous transition. Notice also that the Heisenberg point $\theta=0$ is a transition point for SO(4), which is quite different from the SO(3) (Haldane) and SO(5) (dimerized) cases.

\subsection{Discussion and arbitrary-$n$ behavior} Let us now discuss two aspects of the above results. First, notice that all the features observed above for SO(4) and SO(5) are also common to our SO(3) calculations. Thus, we conjecture that these are \emph{generic properties of the many-body entanglement of the system regardless of the value of $n$}. Second, we believe that there are good chances that $\theta=3\pi/2$ \emph{is a point of infinite entanglement length and finite correlation length for any $n$}.

Let us give an intuitive argument in favor of this. First, we remind that at this point the ground state subspace is made of two degenerate dimerized states, each one of them adiabatically connected (and thus with the same long-range properties) to a dimerized state of SU($n$) singlets between nearest neighbors, see Ref.~\cite{bethe-1, bethe-2, bethe-3} for a proof \footnote{These singlets can be interpreted as having SO($n$) symmetry, but acquire SU($n$) symmetry at this point \cite{bethe-1, bethe-2, bethe-3}.}. These dimerized states are known to have infinite localizable entanglement (and thus infinite entanglement length) if measurements are allowed on pairs of nearest-neighbor spins that do not share a singlet. Such long-range properties should then also belong to the two dimerized ground states at $\theta=3\pi/2$. Also, since this point in the parameter space belongs to a dimerized gapped phase, translational invariance is broken in the ground state of the system towards one of these two dimerized states. Therefore, the system should display an infinite entanglement length whereas the correlation length remains finite because the whole dimerized phase is gapped. This, in fact, is very similar to what happens in the ground state of the deformed AKLT model \cite{geelusive, localizable} and two other MPS models (see Ref.\cite{geqpt-2}). A quantitative analysis of the entanglement length would be needed to ascertain this intuitive argument. Such an analysis, though, is currently beyond our reach.

\subsection{Comparison to other approaches} A comparison between our results and those obtained by other methods is in order. It is known that several quantities may be singular at the well-known transition points of the model, but not around other possible transitions. For concreteness, we consider here the case of SO(3), and calculations of the mutual information \cite{mutual}, fidelity susceptibility \cite{fs}, fidelity diagram, R\'enyi entropies and correlation functions. Specifically: (i) The mutual information was computed in Ref.~\cite{mutual} for finite systems up to 14 sites using exact diagonalization, and while it detects some of the transitions in the system, its behavior is analytic around $\theta = \pi/4$ and $7 \pi/4$, as well as around the conjectured nematic phase at $\theta = 3 \pi / 2$. (ii) In Ref.~\cite{fs} the authors perform a study of the fidelity susceptibility up to 12 sites using exact diagonalization, and no anomalous behavior is observed at the KT transition at $\theta = \pi / 4$ and at the second order transition at $\theta = 7 \pi / 4$. Nothing special is either observed around $\theta \approx 1.34 \pi$ and $\theta = 3 \pi / 2$. This fact is interesting since the fidelity susceptibility is, essentially, the curvature in the axis direction along a diagonal of the fidelity diagram. Its magnitude seems then not to be sufficiently strong to identify certain non-analiticities in the system that can otherwise be detected by looking at the whole picture of this diagram (see supplementary material). (iii) We have compared our results to those obtained by analyzing the R\'enyi entropies of half an infinite chain, which include the von Neumann entropy and the single copy entanglement as particular cases \cite{entropysingle-1, entropysingle-2, entropysingle-3, entropysingle-4, entropysingle-5}. The R\'enyi entropies and, in fact, the entanglement spectrum of half an infinite chain \cite{entspec-1, entspec-2, entspec-3}, are a direct byproduct of the iTEBD method that we employed to approximate the ground state wavefunction. We have seen no signatures of anomalous behaviors in these quantities at $\theta = 3 \pi / 2$. (iv) Finally, we also computed several two-point correlation functions from the obtained MPS of the ground state, and saw no clear sign of anomaly around the integrable point $\theta = 3 \pi/2$. All these observations are in contrast with the behavior of the GE.

\section{Conclusions} In conclusion, we have investigated the phase diagram of the SO($n$) bilinear-biquadratic quantum spin chain using the GE, and by considering numerically the cases of $n=3,4$ and $5$. We have conjectured that our numerical observations are also valid for arbitrary SO($n$). Furthermore, we have seen that the GE provides a remarkably rich behavior in the phase diagram as compared to other quantities.

To finish, let us mention that it would be interesting to determine which quantities, apart from the GE, can be useful to determine the rich phase diagram of the models considered here. For instance, a detailed analysis of the degeneracies in the entanglement spectrum of the ground state  for arbitrary SO($n$) should be helpful in determining the relevant phases \cite{es-1,es-2}. This will be the subject of future investigations.

\acknowledgements

R. O. acknowledges ARC, UQ and EU. T.-C. W. acknowledges NSERC and MITACS. Discussions with H.-Q. Zhou and I. Affleck are acknowledged.

\appendix

\section{fidelity diagram for SO(3)}

\begin{figure}
\centerline{\includegraphics[width=9cm,angle=0]{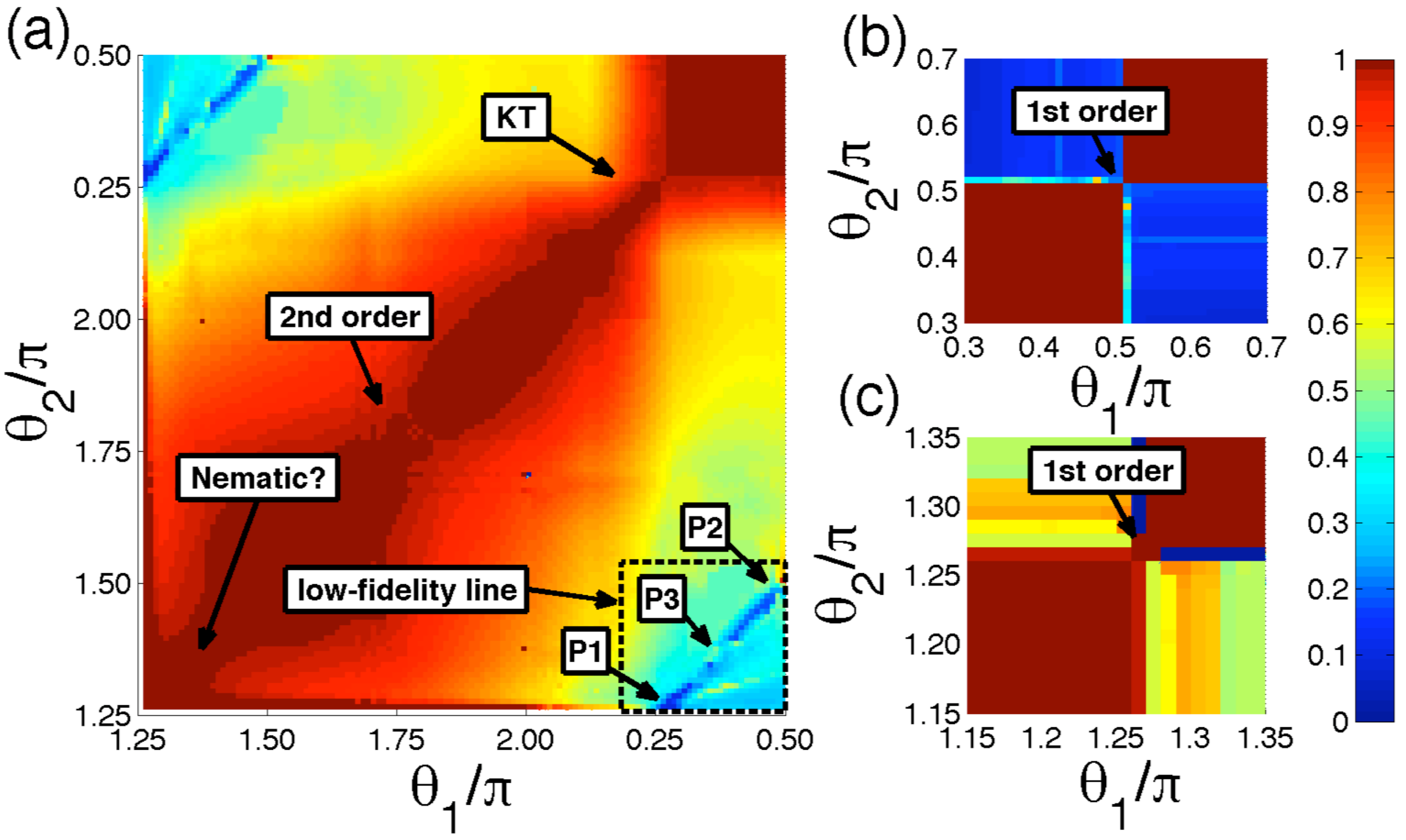}}
\caption{(Color online) Different sections of the FD of the spin-1 bilinear-biquadratic spin chain, capturing a variety of different behaviors: (a) Region $\theta \in (5 \pi/4, \pi/2)$; (b) Region $\theta \in (0.3 \pi, 0.7 \pi)$; (c) Region $\theta \in (1.15 \pi, 1.35 \pi)$.}
\label{fig:fid}
\end{figure}

Let us consider the fidelity diagram (FD) of the system for the case of SO(3). The aerial view of the $(\theta_1,\theta_2)$ plane is represented in Fig.~\ref{fig:fid} (notice that the FD is symmetric). As can be seen in the plot, the FD is discontinuous at $(\theta_1,\theta_2)=(\pi/2, \pi/2)$ and $(5 \pi/4, 5\pi/4)$, in accordance with the first order transitions present at these points. The results indicate that the transition at $(\pi/2, \pi/2)$ seems to be caused by a much stronger discontinuity than the one at $(5 \pi/4, 5\pi/4)$. Furthermore, a pinch point is present at $(\pi/4,\pi/4)$ corresponding to the KT transition in the system. Therefore, these results show that the FD is able to capture the presence of this transition, unlike other closely-related quantities such as the fidelity susceptibility. The presence of another pinch point at $(7 \pi/4, 7\pi/4)$ is compatible with the second order quantum phase transition here.

Importantly, we see a number of other features in the FD. First, we observe a tiny pinch point around $\approx (1.34 \pi, 1.34 \pi) $. This would be compatible with a transition towards a spin nematic phase at this point, yet we stress that we can not conclude on the existence of such a transition based on these results only. Second, we observe a line of low fidelity between the points $P1 = (\pi/4,5\pi/4)$ and $P2 = (\pi/2, 3 \pi/2)$ in the lower half of the plane (and its symmetric counterpart in the upper half plane). There are two remarkable facts about this line. First, notice that it goes between points that are known to be special in the model: at $\theta = \pi/4$ there is a KT transition, at $\theta = \pi /2$ and $\theta = 5 \pi /4$ there are first order transitions, and at $\theta = 3 \pi/2$ the model is exactly solvable. Intriguingly, this last one is the parameter value at which the GE shows a cusp, see Fig.~\ref{fig:GEZoom}(e). Thus, it looks like $\theta = 3 \pi/2$ is also special from the point of view of the FD. Second, the line of low fidelity is slightly interrupted around $P3 \approx (1.34 \pi, 0.37 \pi)$. This is another indication that $\theta \approx 1.34$ is special for the FD (notice also that $\theta = 0.37 \pi$ is not special: it is simply the $y$-coordinate over the line between $P1$ and $P2$ when the $x$-coordinate is $\sim 1.34$). This behavior is also in accordance with our previous observations using the GE.

\end{document}